# Coupling physical understanding and statistical modeling to estimate ice jam flood frequency in the northern Peace-Athabasca Delta under climate change

Jonathan R. Lamontagne[1*], Martin Jasek[2], Jared D. Smith[3]

1. Department of Civil & Environmental Engineering, Tufts University, Medford, MA, USA
2. BC Hydro, Burnaby, BC, Canada
3. Department of Engineering Systems and Environment, University of Virginia, Charlottesville, VA, USA
   * Corresponding Author

Abstract:
The Peace-Athabasca Delta (PAD) of northwestern Alberta is one of the largest inland freshwater deltas in the world, laying at the confluence of the Peace and Athabasca Rivers.  The PAD is internationally recognized as a having unique ecological and cultural significance, and periodic ice jam flooding from both rivers is an important feature of its current ecology.  Past studies have debated whether a change in ice jam flood frequency on the Peace River has recently occurred, and what factors might be driving any perceived changes.  This study contributes to this debate by addressing two questions: (1) what factors are most predictive of Peace River ice jam flooding, and (2) how might climate change impact ice jam flood frequency?  This work starts with a physically-based conceptual model of the necessary conditions for a large Peace River ice jam flood, and the factors that indicate whether those conditions are met.  Logistic regression is applied to the historical flood record to determine which combination of hydroclimatic and riverine factors best predict ice jam floods and the uncertainty in those relationships given the available data.  Winter precipitation and temperature are most predictive of Peace River ice jam floods, while freeze-up elevation contains little predictive power and is not closely related to ice jam flood occurrence.  The best logistic regression model is then forced with downscaled climate change scenarios from multiple climate models to project ice jam flood frequency for a variety of plausible futures.  Parametric uncertainty in the best logistic regression model is propagated into the projections using a parametric bootstrap to sample many plausible statistical models. Although there is variability across emissions scenarios and climate models, all projections indicate that the frequency of Peace River ice jam floods is likely to decrease substantially in the coming decades, and that average waiting times between future ice jam floods will likely surpass recent experience.

## Graphical Abstract:

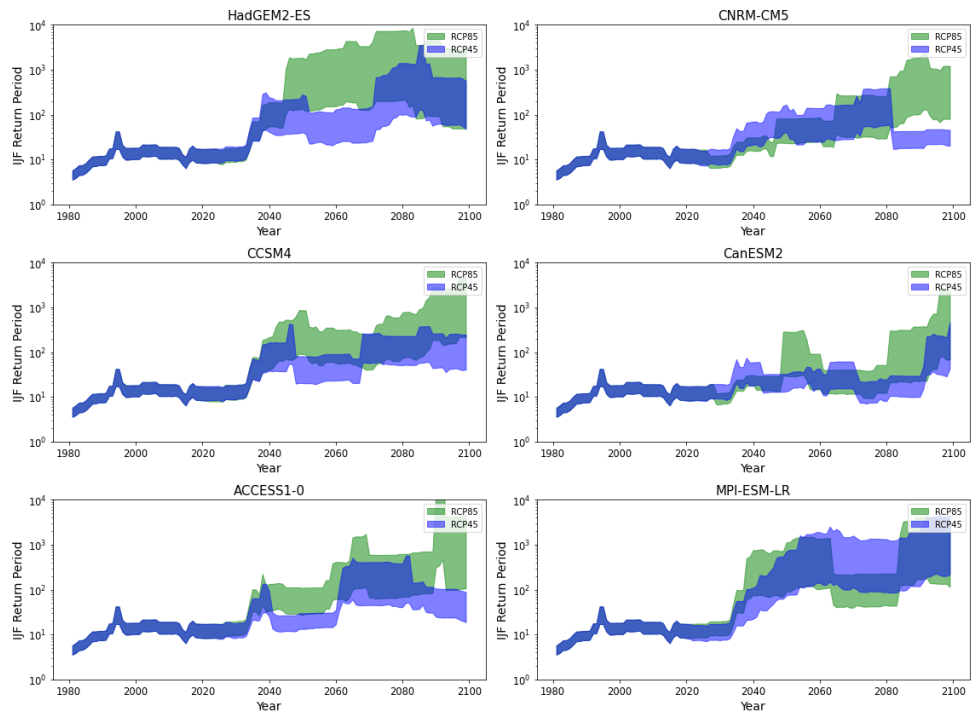

# 1 Introduction

The Peace-Athabasca Delta (PAD, Figure 1) of northwestern Alberta is one of the largest inland freshwater deltas in the world, lying at the confluence of the Peace and Athabasca Rivers. The PAD consists of a delta plain interspersed with many channels, small lakes, and basins that generally flow north into Lakes Claire and Athabasca before draining into the Slave River. The hydrology of the

PAD is complex and during flood times the direction of flow in many channels can reverse and much of the PAD may become inundated. This is most common when large ice jams on the Peace, Athabasca, or Slave Rivers form temporary ice dams during the spring freshet that can result in flow reversals and widespread flooding. Large ice jam floods have presumed biological significance in the PAD, as many restricted (perched) basins (small lakes) only receive inflows during these events (Peterson, 1992; Prowse and Lalonde, 1996; Timoney, 2002). A change in the frequency of such recharge events could affect the long-term hydro-ecological functioning of the PAD in ways that are difficult to anticipate.

The PAD is recognized as having unique ecological and cultural significance: it is within Wood Buffalo National Park, is a Ramsar-Convention Wetland of International Importance, and is part of a UNESCO World Heritage Site. The PAD is home to large populations of muskrat, beaver, herds of free-ranging wood bison, and waterfowl, and it is important to migratory birds, such as the endangered whooping crane. Several Indigenous Nations, including the Mikisew Cree, Athabasca Chipewyan, and Metis rely on resources in the PAD.

The PAD is constantly evolving, with dynamic geomorphology and ecology due to sediment deposition, climate non-stationarity, and other factors. This is especially true of the Athabasca (southern) portion of the delta, which receives substantial sediment loads from the Athabasca River. Starting in the early 1970s, advocates and scientists began to perceive changes in various features of the PAD (Peterson, 1992), including a perceived change in ice jam flood frequency. Since then, a robust debate in the scientific literature has ensued as to whether recent changes are different than historical changes, or consistent with long-term trends or patterns found in archival and paleo-records (Wolfe *et al.*, 2006, 2012, 2020). For example, the apparent reduction in ice jam frequency since the early 1970s roughly coincides with the construction of large dams and flow regulation on the Peace River. There is however, disagreement in the scientific literature about whether, when taking a long-term view, a significant reduction in ice jam frequency has really occurred (Beltaos, 2017; Hall, Wolfe and Wiklund, 2018; Timoney *et al.*, 2018), and what the key predictors of ice jam formation and PAD flooding are (Timoney *et al.*, 2018; Jasek, 2019c). A primary concern for the future of the PAD are the potential impacts of climate change (hydrological, ecological, cultural, etc.), that are difficult to anticipate and may include changes to ice jam flood frequency.

This paper contributes to this debate by addressing two research questions:
(1) What factors are most predictive of large ice jam flood formation in the PAD?
(2) How might climate change alter large ice jam flood frequency in the 21$^{st}$ century?

The second question is predicated on the hypothesis that climatic factors are predictive of ice jam formation, so we first establish this hypothesis is consistent with the data by answering question one. To do this we combine physical understanding of the formation of ice jam floods with statistical inference. Specifically, we use logistic regression to determine which factors are most predictive of the occurrence of ice jam floods over the historical period.

The paper is organized as follows: Section 2 provides further background on the PAD, including more discussion of the PAD's geography, a discussion of physical factors affecting ice jam formation, and past studies of ice jam frequency in the PAD. Section 3 describes the stochastic approach to modeling ice jam occurrence adopted in this study, and Section 4 relates the results of its application to the assessment of ice jam flood frequency under several climate futures. Section 5 contains concluding remarks.

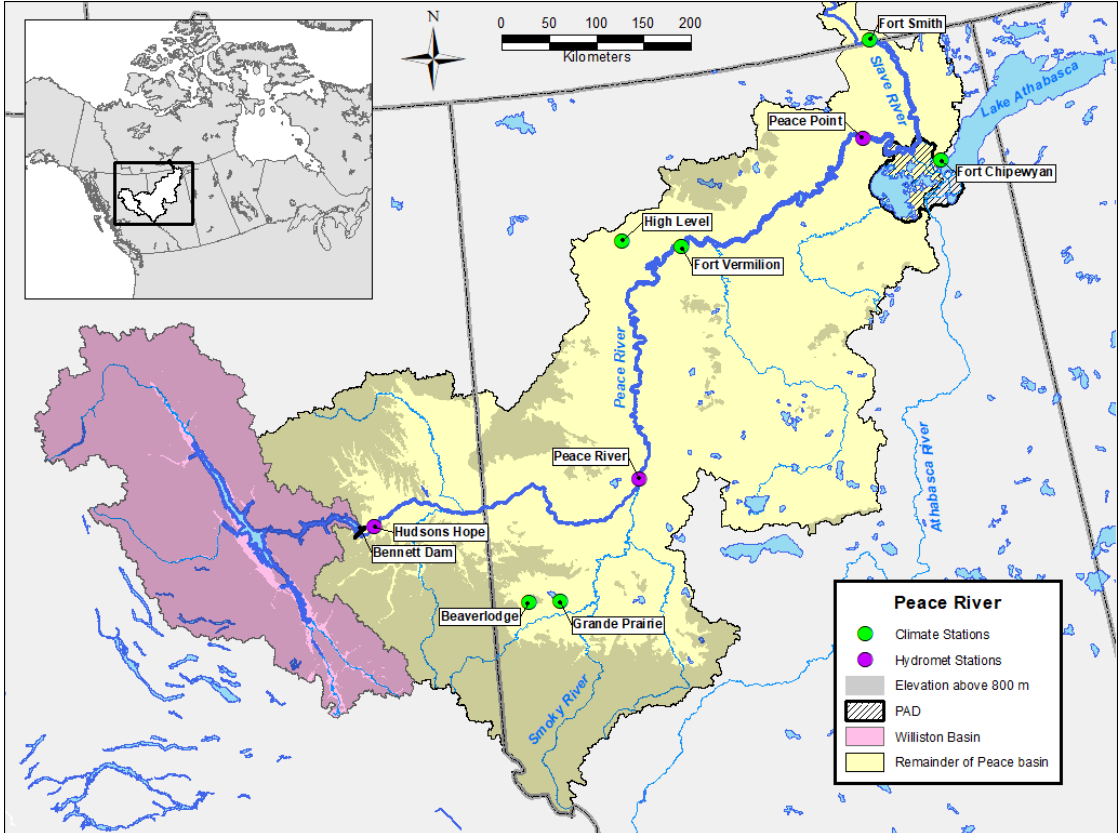

*Figure 1: Map of the Peace-Athabasca Drainage Area and the Peace-Athabasca Delta (PAD). Also shown are meteorological and hydrometric stations from which data were used in the analysis. Most of the snowmelt that drives breakup comes from the lower elevation (yellow) portion of the basin. The pink portion represents the drainage are that is regulated by the Bennet Dam.*

# 2 Background on the Peace-Athabasca Delta (PAD)

This section provides additional background on the PAD's geography, ice jam formation, and previous studies of ice jam frequency. These physical factors of ice jam formation are critically important to inform the selection of variables that explain ice jam flood occurrence and to interpret the statistical models of ice jam flood frequency presented in Section 3.

## 2.1 PAD Geography and Importance of Ice Jam Floods for Basin Recharge

The PAD lies at the confluence of the Peace and Athabasca Rivers in northwestern Alberta and consists of a network of small lakes and basins spread across a large delta plain. The lakes and basins are typically categorized into three types: open, semi-restricted, and restricted. Open basins are mostly large lakes that are directly connected to major rivers and delta channels. Semi-restricted basins are connected intermittently or seasonally during moderately high-water events. Restricted basins typically only receive river water replenishment under very high-water levels. Open water (summer floods) are not capable of generating such water levels in the PAD; ice jams on the Peace, Slave, or the Athabasca Rivers and associated flooding are needed to replenish these restricted basins. Cycles of recharge and desiccation are an important aspect of the ecological community in these basins, and changes to that cycle will affect the ecology in ways that are hard to anticipate (Timoney, 2013).

Both the Peace and Athabasca Rivers play a role in the water balance and evolution of the PAD, the latter entering the PAD from the south and the former running along the northern edge. Much of the concern about the drying of the PAD focuses in the northern portion, which is most closely linked to the Peace River. The Peace River is connected to the PAD through several channels, which generally flow north, out of the PAD into the Peace River but occasionally reverse flow (southward) during high flow events or ice jams on the Peace and Slave Rivers. Overland flow can also occur during major ice jam floods. Interestingly, Timoney [2013] indicates that on average, relatively little water entering the PAD originates from the Peace River, but that high flows on the Peace River play an important role in forming a hydraulic dam that maintains water levels in the northern PAD. In the mid-1970s, low flows on the Peace River prompted the placement of several rock weirs in the Rivière des Rochers, Revillon Coupé and Chenal des Quatre Fourches to maintain higher water levels in the northern PAD during such low-flow events.

The Peace River is partially regulated by the Bennett and Peace Canyon Dams about 1200 km upstream of the PAD in British Columbia. These facilities, along with dams on the Columbia River and other smaller facilities, provide 97.8% of British Columbia's power at low cost and without emitting significant greenhouse gasses compared to fossil fuel energy sources (BC Hydro,

2019).  Flow regulation on the Peace River started in 1972 when the Bennett Dam was completed, though filling of the Williston Reservoir impacted flows from 1968-1971.   Coincidentally, the decade prior to regulation was much wetter than the decade after regulation, thus making it difficult to attribute perceived changes (drying) in the PAD to the dam itself. This is indicated by drying trends in adjacent, unregulated basins that roughly coincide with the completion of the Bennet Dam (Timoney, 2002). The regulation of the Peace River does reduce the magnitude of summer floods because water is stored in the Williston Reservoir for peak power demand in the late fall and winter, though in general summer floods are insufficient to refresh restricted basins in the northern PAD.

Spring flows are not affected to a great degree by regulation of the Williston Reservoir. Most of the Peace River flow during ice breakup comes from the unregulated portion of the Peace basin (see Figure 1) so the regulated flows play a smaller role during the potential ice breakup period in the PAD.  Additionally, flow due to regulation at the time of breakup increases the natural river flow on average and remains unchanged at the time of ice jam PAD flooding (Beltaos and Peters, 2019), therefore, from a purely available discharge perspective, there is no reason to suspect that the initiation of an ice jam flood, and its first week of flooding, is reduced by regulation. To test the impact of regulation at the time of freeze-up, we evaluate regulated river flows and freeze-up as possible predictors of ice jam floods in the statistical analysis presented in Section 3.

## 2.2   Conditions that Favor Ice Jam Formation and Flooding in the PAD

For an ice jam flood to occur in the PAD, a sequence of favorable conditions must develop.  Jasek, [2019ab] developed a conceptual model of these conditions and potential outcomes that is central to the current analysis.  Because many readers will not be familiar with ice jam floods in general, and few with PAD ice jam floods in particular, we begin with an introduction to ice jam floods followed by a narrative of how a large ice jam flood might occur in the PAD.

There are two types of river ice breakups: thermal and dynamic (mechanical) (Beltaos, 2008). The difference between the two has to do with the ice breakup resistance (related to ice strength). Breakup resistance is the ice cover's ability to resist the downstream forces from the water and debris flow, as well as gravitational forces. If breakup resistance is low, a large spring freshet can initiate a dynamic breakup where the ice cover rapidly cracks apart and flows downstream, with the leading edge of the breakup moving quickly down river. On the other hand, if the ice is strong and breakup resistance is high, a thermal breakup can occur wherein the leading edge of the ice cover is able to resist downstream forces and the ice cover melts in place due to thermal sources, such as warm water, air, and solar radiation. Ice jams usually occur due to dynamic breakups, but a dynamic breakup alone is not sufficient to cause a substantial ice jam flood in the PAD: the dynamic breakup may flush through the reach and down the Slave River, such as occurred in 2018 (Jasek 2019a), or be flushed into Lake Athabasca via the Rivière des Rochers, such as occurred in the spring of 2020. The following paragraph illustrates the complexity in physically determining whether or not an ice jam flood will occur in the PAD by presenting sequences of favorable conditions that must develop.

For an ice jam to form, there must first be a large snowpack in the tributary basins of the Peace River, most notably the Smoky River Basin (Figure 1). If the spring warm-up is rapid and sustained, the resulting freshet can dislodge the river ice cover in the tributaries. As ice conditions, warming, and channel hydraulics are not uniform in the tributaries, sections of continuous ice cover and open flow emerge, and jams begin to form. If the freshet is strong and sustained, these jams move through the tributary network, eventually entering the main-stem of the Peace River, which can initiate a dynamic breakup of the Peace River ice cover. As in the tributaries, the ice resistance along the Peace River is highly variable, and sections of open flow, moving ice, and jams can begin to form as long as flows remain sufficiently high (high snowpack, sustained warming). Under these conditions, a large ice jam can begin to form along the leading edge of the breakup. That jam can be dozens of kms long and impound enormous volumes of water. When that jam releases it is akin to a dam failure, releasing the impounded water and triggering breakup of the ice cover for hundreds of kms [Jasek, 2005, 2017]. At this point, if the leading edge of the ice breakup encounters a stretch of highly resistant strong river ice, the breakup can stall long enough for the impounded water to bleed away under the remaining solid ice cover. If the ice jam remains in place long enough, this bleeding will either reduce the volume of the ice jam flood arriving at the PAD or prevent the flood altogether. However, if the down-stream Peace River ice resistance is low enough, the cascading breakup will continue more or less unabated until it reaches the Upper Slave River and PAD region. If the ice resistance in the PAD is low, the cascading breakup may flush through the upper Slave River and produce only a brief period of moderately high water and no major PAD flood will occur. If the ice resistance in the PAD is high, the cascading breakup will arrest and there is a potential for a large PAD ice jam flood.

The complexity described in this narrative leads to significant uncertainties: quantifying ice resistance over hundreds of kms of river alone is challenging. However, it is possible to construct a conceptual model of breakup factors [Jasek, 2019ab] that we divide into four categories:

1. Discharge potential for breakup: snowpack and Peace Canyon Releases
2. Sustained warm-up: duration and rapidity of spring thaw
3. Breakup ice resistance upstream of the PAD
4. Breakup ice resistance in the PAD and upper Slave River

For illustrative purposes we imagine that each factor is a binary variable ("high" or "low"), though in reality each factor (if it were possible to measure) is continuous. There are then 16 possible combinations, as summarized by Jasek [2019c], of which the six most likely to result in a PAD ice jam flood are reported in Figure 2. Based on historical observations, only one of the 16 combinations is expected to have a high likelihood of generating a large ice jam flood (combination

3). Combination 1 has the possibility of a range of outcomes that includes the potential for a large ice jam flood in the PAD, but has a smaller expected likelihood than combination 3. The other 14 combinations result either in a thermal breakup, a dynamic breakup flush-through, or a small or moderate ice jam flood in the PAD.

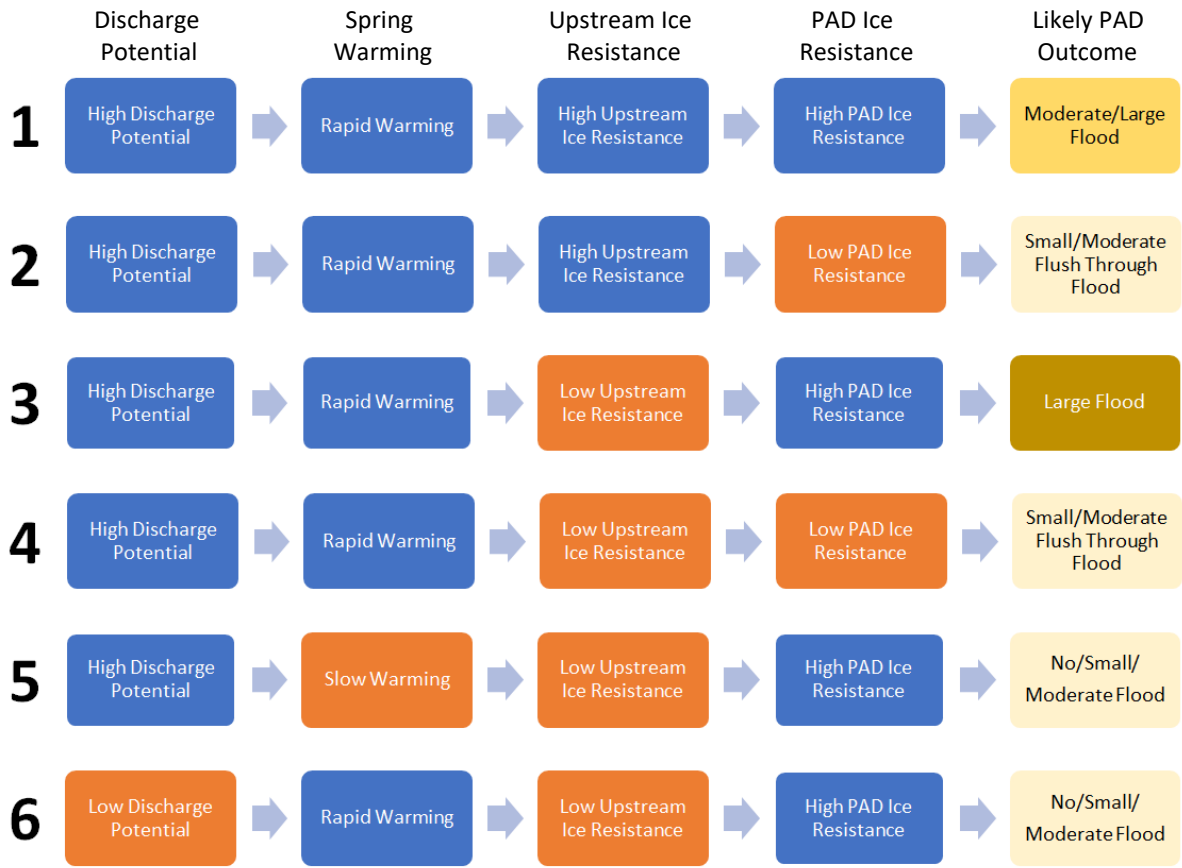

*Figure 2: Six combinations of ice breakup factors from Jasek (2019c) that can lead to dynamic breakups at the PAD. In 1 whether a large or moderate flood occurs depends where the dynamic breakup stops to form an ice jam, as there is high ice resistance both upstream and at the PAD. "Flush throughs" (2 and 4) are dynamic breakups that continue past the PAD reach and result in no or moderate/short-lived flooding because an ice jam does not form. Conditions in 3 lead to the highest probability of a large ice jam flood. "No Flood" noted in 5 and 6 indicates that a thermal breakup is still possible in these scenarios but a dynamic breakup that causes a small or moderate flood is also possible. The remaining 10 combinations (not shown) result in no floods or thermal breakups. A full set of 16 combinations and their outcomes in flow chart format is shown in the supplementary materials.*

The first breakup factor is discharge potential, which has two components: regulated Peace River flows (from the Peace Canyon and Bennett dams) and unregulated inflows from tributaries

(spring freshet). At the time of breakup, unregulated inflows from snowmelt make up the overwhelming majority of Peace River flows (about 75-90%) as a result of the elevation topography where the lower elevation snow which is in the larger and unregulated portion of the basin (Figure 1). Thus, the accumulated snowpack in the unregulated tributary basins of the Peace River is the leading factor of discharge potential. The snowpack must be sufficient to produce the high flows needed to initiate and sustain a dynamic breakup that can lead to large ice jams.

Historical snow survey data is limited, so cumulative winter precipitation at Grande Prairie (in the Smokey River Basin) was used as a proxy for total winter snowfall that was available at freshet to drive break-up. Missing data at Grande Prairie was filled in using the record from Beaverlodge, which is 36 km away and at a similar elevation (see Figure 1). Initially, cumulative winter precipitation was computed from November to April, which is reasonable for in that region over the historical period as rainfall consistently changes to snowfall in early November. However, this assumption presents a problem when considering future climate change (Section 3.3), as the rainfall season is likely to extend, necessitating a different approach. For each year, it was assumed that precipitation would accumulate as snow until the spring once sustained degree-days freezing (DDF) were negative; that is if the cumulative DDF became negative but eventually returned to zero in the early winter, it was assumed that all the snow that had fallen to date had melted and the accumulated precipitation was zeroed out and began to accumulate again. Over the historical period this approach made only minor changes to the first approach, but it made a substantial difference for future (warmer) scenarios. Mid-winter warm spells were assumed to not affect the total amount of water available at break-up because the ground would be frozen, and the meltwater would freeze to the ground surface and be available in the spring for melt.

The second breakup factor is the springtime warming. A rapid warming can result in considerable snowmelt before the river ice has deteriorated, leading to favorable conditions for ice jam formation and dynamic breakups upstream of the PAD. On the other hand, slow warming allows time for the river ice to deteriorate to the point that any potential dynamic breakup is likely to flush through and not jam. If warming is slow, it is also possible the breakup will be thermal and not produce the high-water levels needed to recharge restricted basins in the PAD. Without rapid springtime warning, it is unlikely that a PAD ice jam flood could occur.

To quantify the rapidity of warming, various statistics based on the cumulative degree-C-days of thaw (DDT) were tried and referred to as "Melt Tests." This was based on an examination of known dynamic break-up dates at the Town of Peace River and the PAD, and the corresponding DDT's at Grande Prairie. The former ranged from 33 to 65 DDT and the latter ranged from 65 to 139 DDT. The important dynamic to capture is not the accumulation of DDTs, but the speed of the accumulation (rapidity of the thaw). Various "Melt Test" statistics were tried and the most successful one is presented in this paper: the number of days to accumulate from 40 DDT to 150 DDT. The smaller this number the more rapid the warming. The value of 40 DDT is enough to get a dynamic break-up of the Smoky River going and the remaining DDT to 150 is enough to sustain the high freshet flows on the Peace River to keep the break-up predominantly dynamic all the way to the PAD and ultimately produce enough volume of water to flood the PAD.

The final two breakup factors relate to upstream and downstream ice breakup resistance. Breakup resistance plays a key role in determining where ice jams form, how long they remain and whether a dynamic breakup is initiated. Breakup resistance is related to ice strength and depends on the failure mode (e.g. bending, compression, tension, shear). It is not uniform along the river channel, making it difficult to quantify or measure directly over large regions like the PAD. Nevertheless, high ice resistance in the PAD is critical for a large ice jam flood in the PAD, as it ensures the ice is strong enough to resist a flush-through event. Large ice jam floods could happen with either low or high ice resistance upstream of the PAD, but they are more likely with low upstream ice resistance as this encourages a dynamic breakup on the Peace River. Many meteorological parameters determine ice break-up resistance (Table 1). Here-in, the most important (DDF) was considered as it determines the thermal ice growth and ultimate ice thickness before break-up. For downstream ice resistance, Fort Chipewyan temperatures were used with missing data determined by Fort Smith temperatures (Figure 1) adjusted by an average monthly offset determined by coincident data between the two stations. Similarly, for upstream breakup resistance, this was done with Fort Vermilion temperature data, where missing data was determined from High Level.

*Table 1: Factors that affect ice breakup resistance.*

| Factors Affecting Ice Resistance | High Ice Resistance | Low Ice Resistance |
|---|---|---|
| Freeze-up Elevation | High, wider ice sheet that takes more downstream force to push around river bends and through constrictions | Low, narrower ice sheet that takes less downstream force to push around bends and through constrictions |
| Total Degree-Days Freezing (DDF) in winter | High, builds thick ice | Low, leads to thin ice |
| Degree-days of thaw prior to discharge increase | Low, less ice melt and strength decay so ice remains strong | High, more ice melt and strength decay so ice is weakened |
| Snowfall on ice in early winter | Low, less insulation on ice cover allows the cold temperature to grow the ice thickness | High, insulates ice cover from cold air temperatures, reducing the rate of ice growth |
| Snowfall on ice in late winter | High, protects ice from solar radiation that reduces the rate of ice strength decline and melt | Low, reduces protection from solar radiation, allowing for ice strength deterioration and thinning |
| Formation of a snow-ice layer | Present, protects ice from solar radiation that reduces the rate of ice strength decline and melt | Absent, reduces protection from solar radiation, allowing for ice strength deterioration and thinning |
| Ice cover formation: frozen frazil vs. clear ice | Frazil, provides protection from solar radiation that reduces the rate of ice strength decline and melt | Clear Ice, reduces protection from solar radiation, allowing for ice strength deterioration and thinning |
| Cloudy vs. Sunny Skies | Cloudy, less solar radiation that reduces ice strength deterioration and melt | Sunny, more solar radiation causing rapid ice strength deterioration and melt |

Figure 3 plots four factors that should, according to the conceptual model in Figure 2 lead to large ice jam floods. Years in which a large ice jam flood occurred are plotted in orange. In each panel, the vertical axis is Grande Prairie/Beaverlodge Winter Precipitation, which is a measure of discharge potential. Fort Chipewyan and Fort Vermillion DDF are measures of PAD and upstream ice resistance, respectively. Freeze-up elevation is also widely considered an important factor for ice resistance, and is plotted in the lower right panel. Melt Test is a measure of the rapidity of spring warming and is plotted in the lower right. Other variables that were considered are reported in Table S. 1.

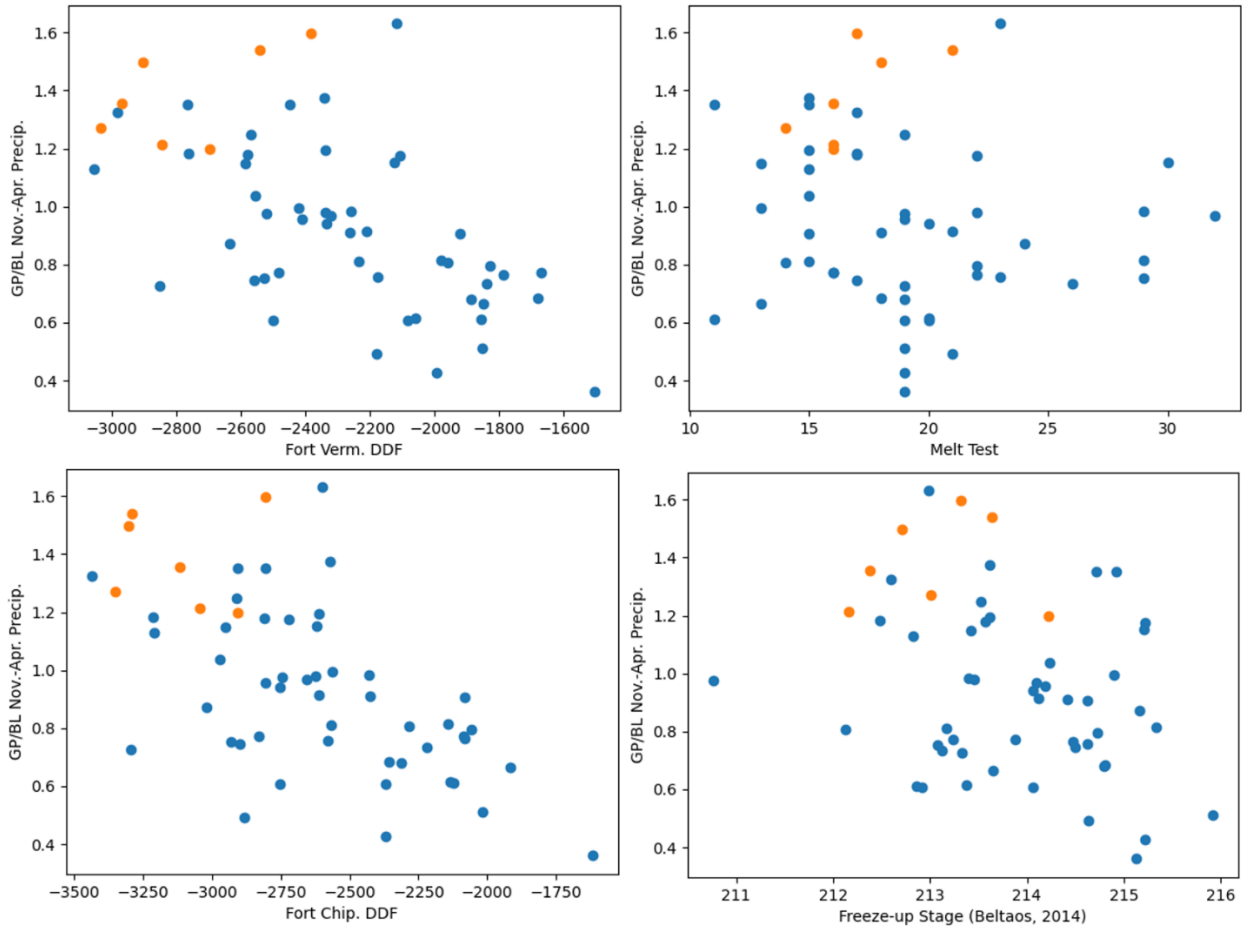

*Figure 3: Exploratory data analysis of several factors and their relation to ice jam flood occurrence (1962-2020). The y-axis is the same for all four panels and represents the total Nov.-Apr. precipitation at Grande Prairie and Beaverlodge. The remaining factors are: DDF at Fort Vermillion (top left) and Fort Chipewyan (bottom left), freeze-up elevation (bottom right), and melt test (top right).*

From Figure 3, we note that large ice jam floods tend to occur when precipitation is higher than average and temperature is lower than average, though not all points yin that plot region are flood years. When precipitation is high, years with lower melt test and freeze-up elevations also tend to result in large ice jam floods, though these relationships seem less clear. Some previous studies (Prowse and Conly, 1998; Beltaos, Prowse and Carter, 2006) have have hypothesized that higher post-regulation freeze-up elevations have increased breakup resistance in the spring, which has in turn prevented dynamic breakups from occurring (Beltaos, Prowse and Carter, 2006). It could alternatively be argued that a higher freeze-up elevation is needed in the PAD reach to arrest dynamic breakups, as a low freeze-up elevation is required upstream of the PAD to allow for a dynamic breakup to initiate. It's a delicate balance because the winter and spring weather are similar a few hundred km away and thus correlated. The significance of freeze-up elevation and these other factors on predictions of large floods is examined in Section 3.

## 2.3   Past Studies of Ice Jam Frequency in the PAD

In the engineering and scientific literature, ice jam frequency analyses can take on three forms, roughly distinguished by their objectives, though each are interconnected. First, frequency-stage or frequency-flow analyses aim to provide planners with flood risk estimates, typically in a relatively narrowly defined spatial domain (USACE, 2011; Ahopelto *et al.*, 2015). Those studies are often concerned with design flows of all types, but in ice-prone regions the largest floods are often the result of ice jams. Consequently, knowing the frequency of such jams is critical. The second type of ice jam frequency analysis focuses specifically on the frequency of ice jam occurrences, or occurrences that exceed some threshold (Timoney *et al.*, 1997; Beltaos, 2014). Those analyses focus on the occurrence of the flood and not the magnitude once it occurs (Timoney *et al.*, 2018). A third type of ice jam frequency study focuses on the frequency with respect to both arrival and magnitude of ice jam floods (Lindenschmidt *et al.*, 2016), including under various climate change scenarios

(Das, Rokaya and Lindenschmidt, 2020). This paper focuses squarely on the second two types: the arrivals of ice jam floods and the arrival rate may be impacted by climate change.

Quantifying the historical ice jam flood frequency in the PAD is exceptionally challenging due to the dearth of available data, both over time and space. The systematic hydrologic record in the PAD dates only to 1972 and includes only five ice jam floods (all post-regulation), so any statistical analyses based on these data alone will be imprecise. The immense size of the PAD, its low population density, and dynamic nature complicate the interpretation of pre-record data sources. Observing an ice jam flood in the PAD is much more complicated than observing a flood at a specific point or river reach. Depending on magnitude, the inundated area may vary by thousands of square kilometers, so that many ice jam floods may have occurred without being observed. When ice jam floods or large inundations were observed and recorded, the nature of those records make it difficult to infer their magnitude and, in some cases, the exact date.

There are essentially two approaches to characterizing historical ice jam flood frequency: analysis of historical records (Timoney, 2009) and paleolimnology (Wolfe *et al.*, 2012, 2020). Using historical written texts, it is possible to reconstruct historical flood records to the late 19$^{th}$ century (Timoney, 2009), while paleolimnology can provide data back several centuries to a millennium (Smith, 2003; Hugenholtz, Smith and Livingston, 2009; Timoney, 2013). The issue with the former is that our certainty concerning the exact date and magnitude of a flood diminishes the further back in time we consider. Timoney [2009] ascribes magnitudes dating to the late 19$^{th}$ century, but these contain some level of subjectivity and floods may have been missed (Wolfe *et al.*, 2020). The issue with paleolimnological reconstructions is that they lack precise temporal resolution, and the spatial resolution is fairly sparse. So, we can be fairly sure that we've captured floods that have occurred in a place in the PAD, but we cannot be sure precisely when they occurred or the extent of inundation. Still, from both sources two things are clear: the frequency of ice jam floods seems to have been non-stationary for at least a few centuries with average return periods ranging from 6 to 25 years, but for the last 200 years there has been a flood on average once every six years (Timoney, 2013).

Since the Bennet Dam's construction, much of the scientific focus on PAD ice jam flood frequency has been on the seeming decline in large ice jam floods and the coincident reduction in PAD water levels. Much of this debate (Beltaos, 2017, 2019; Jasek, 2017; Hall, Wolfe and Wiklund, 2018; Timoney *et al.*, 2018) has focused on two questions; (1) has there been an actual decline or is it simply random variability, and (2) if there has been a decline, what is the cause? Previous studies have hypothesized that higher post-regulation winter flows have resulted in higher freeze-up elevations that have had a deleterious impact on ice jam flood frequency (Prowse and Conly, 1998; Beltaos, Prowse and Carter, 2006; Beltaos, 2014, 2017). Beltaos [2017] uses statistical methods to demonstrate declines in frequency and attribute this in part to regulation by the Bennet dam. Those results are reported with questionable degrees of certainty (Hall, Wolfe and Wiklund, 2018; Timoney *et al.*, 2018). Others have shown that the perceived changes in ice jam frequency are not statistically significant, which is simply to say that there is insufficient data to say whether a decline in ice jam frequency has occurred (Timoney, 2013; Hall, Wolfe and Wiklund, 2018; Timoney *et al.*, 2018), so that any discussion of precise attribution to regulation seems premature. To explore whether regulated winter flows are predictive of ice jam floods, early winter flows at three points on the Peace River having varying degrees of regulation are considered in the analysis in Section 3.

In studying attribution, Beltaos [2014] raises an interesting question as to the leading drivers of ice jam flood formation. Many studies cite freeze-up elevation as a leading cause (Prowse and Conly, 1998; Beltaos, 2014, 2017), for which there is some theoretical support (see Section 2.2). The first research question in this paper tests whether the data in the PAD is consistent with this hypothesis using statistical regressions to identify the leading factors in ice jam flood formation. The next question asks how ice jam flood frequency in the PAD may change under climate change.

Others have considered this question for various reaches around the world (Ahopelto *et al.*, 2015; Lindenschmidt *et al.*, 2016; Das, Rokaya and Lindenschmidt, 2020). Particularly relevant to this work are recent developments by Lindenschmidt *et al.* [2016], who couple hydrodynamic and river ice models to construct a frequency-stage distribution for the Town of Peace River while accounting for uncertain in parameter estimates within a Monte Carlo framework Das, Rokaya and Lindenschmidt [2020] show that under various climate impacts, ice jam frequency is likely to decrease on average, though very large ice jam flood events may persist. This is promising work that should be pursued further, though it is data and computationally intensive. We propose a complementary, purely statistical approach, described in the next section.

# 3  Modeling Ice Jam Flood Frequency using Logistic Regression

The occurrence of an ice jam flood is a stochastic process. That is, we cannot deterministically predict an ice jam flood because (1) we do not fully understand ice jam dynamics, and (2) even if we did there are determining factors that are unobservable, either practically or technically, so that the occurrence of an ice jam flood is random. Indeed, the 2020 breakup provides a marvelous example wherein all signs pointed to a flood, but even to the last moment it was unclear whether a major jam would or would not form. This then casts considerable doubt on the whole endeavor of deterministically projecting ice jam floods over an extended period using climate projections: if we cannot be sure of a jam in a few days, how sure could we be in 30 or 100 years? Furthermore, whether an ice jam happens to occur (either in history or in a simulation) is little more than an anecdote. History is one realization of a stochastic process that will never be repeated; Global Climate Model (GCM) simulations are each one realization of a model that will never exactly occur. Relying solely on these is like relying on a single spin of the roulette wheel to infer the long-term prospects of the gambler. If the chances of winning are low (as are the chances of an ice jam flood), then relying on a dozen spins, or a dozen model realizations, is hardly better.

What is needed to evaluate the long-term prospects of ice jam floods in the PAD is a stochastic approach based on probability, which after all was invented to quantify the risk in games of chance.

We begin by modeling the occurrence of an ice jam flood using a Bernoulli random variable, $I$, that takes a value of 1 if a flood occurs and 0 if one does not occur. The distribution of $I$ is summarized by a probability mass function of the form:

$$P_I(i) = \begin{cases} p_{flood}, & \text{if } i \text{ is } 1 \\ 1 - p_{flood}, & \text{if } i \text{ is } 0 \end{cases}$$

*1*

Thus, in any given year the probability of an ice jam flood occurring is $p_{flood}$. One estimate of $p_{flood}$ could be obtained by counting the number of historical floods and dividing by the number of years of record:

$$\hat{p}_{flood} = \frac{N_{flood}}{N}$$

*2*

A distinction should be made here between $p_{flood}$ and $\hat{p}_{flood}$, where the former is the true probability of a flood (theoretical) and the latter is an estimate from data (empirical). This is an important distinction, as we can never know $p_{flood}$. The best we can hope for is to improve our estimate, $\hat{p}_{flood}$. Scientists may endeavor to do this with longer record lengths, or with better estimation techniques than equation 2. Still, no matter how sophisticated the method, $\hat{p}_{flood}$ will never be perfect and should not be confused with $p_{flood}$.

The keen reader will no doubt be asking whether a single $p_{flood}$ is of much use, and whether the stationary distribution in equation 1 is sensible. Leaving aside, for a moment, long-term changes to the PAD, we know that certain conditions favor ice jam floods. That is to say, our mechanistic understanding of ice jam floods suggests that the likelihood of an ice jam flood changes from year to

year based on hydroclimatic or other factors, $X$. This can be represented using conditional probability:

$$\Pr(I = 1 | X = x) \tag{3}$$

This is read as the "probability that $I = 1$, given that factor $X = x$." It is possible to condition the probability of an ice jam flood on more than one hydroclimatic variable, in which case $x$ would be a vector and written in bold, $\boldsymbol{x}$. If $x$ bears no relation to the occurrence of ice jam floods, the conditional probability should be equal to the unconditional, long-term probability:

$$\Pr(I = 1 | x) = p_{flood}$$

For instance, we would not expect the winning percentage of the Vancouver Canucks ice hockey team to have any impact on the occurrence of an ice jam flood, so we would expect from data:

$$\Pr(I = 1 | Cunucks\ win\ percent) = p_{flood}$$

So, the fortunes of hockey teams likely contain little information about the likelihood of an ice jam flood. In contrast, we expect that discharge potential (defined above) is likely to have a strong relationship with the occurrence of ice jam floods, so one would expect:

$$\Pr(I = 1 | High\ Discharge\ Potential) > p_{flood}$$

and

$$\Pr(I = 1 | Low\ Discharge\ Potential) < p_{flood}$$

Thus, our theory suggests that one could refine the probability of an ice jam flood in a given year if one had information about the discharge potential. This theory could be tested with data, as discussed below, but this raises two issues. First, in testing this theory by fitting models to data, our estimates will have error, and in general $\Pr(I = 1 | x) \neq \widehat{\Pr}(I = 1 | x)$, so one must be cognizant of the precision of any fitted model. Second, one must recall the old adage that "correlation is not

causation," and view any empirically developed relationship with due skepticism. So, for instance, one may find that ice jam floods are more likely to occur in years with fewer muskrat litters. This is hardly evidence that the muskrats are causing the ice jam floods; it could be that conditions favoring muskrats also favor ice jams, or it could be chance. If the objective is only to predict the probability of an ice jam flood, one may not care. If it is to gain understanding of the underlying system, one may care.

Beltaos [2014] used conditional probabilities to attribute the perceived changes in ice jam floods to different sources. In doing so, that work forwarded an interesting theory: that the conditional probability of an ice jam flood, given some forcing, is invariant in time. That is $\Pr(I = 1|x)$ is stationary, at least over the timescales relevant to human decision making. Instead, Beltaos [2014] argues that the perceived changes in PAD frequency are due to changes in the forcings, $X$. We can understand this theory through the definition of conditional probability:

$$\Pr(I = 1 \cap x) = Pr(I = 1|x)Pr(x)$$

where $\cap$ is the logical "and" and $\Pr(I = 1 \cap x)$ is read "the probability $I = 1$ (i.e. an ice jam flood) and $x$ both occur." Beltaos, [2014] contends that if there has been a change to the left (the joint occurrence of ice jams and forcing level $x$), it is because there is a systematic change to the forcings $Pr(x)$, not the response to those forcing, $Pr(I = 1|x)$. Given that the response to the forcings is largely a matter of geography, bathymetry, etc., this seems a reasonable assumption over a few decades or a century.

Adopting this assumption allows us to use a stationary model to estimate $\Pr(I = 1|x)$. Specifically, we adopt a logistic regression model. Logistic regression is a standard method used in classification, wherein one attempts to predict the value of a Bernoulli random variable given predictors (Wilks, 2019). Standard linear regression estimates the conditional mean of the dependent variable, given the value of the independent variable(s):

$$E[I|x]$$

Because $I$ is a Bernoulli random variable:

$$E[I|x] = \Pr(I = 1|x)$$

So, one may be tempted to apply linear regression directly to the ice jam flood record. However, there is no guarantee that the fitted model will only return estimates of $\Pr(I = 1|x)$ in the range $(0,1)$, indeed it often will not (Hastie, Tibshirani and Friedman, 2009). Logistic regression accounts for this by adopting the logistic function, which varies between 0 and 1:

$$Pr(I = 1|x) = \frac{\exp(\boldsymbol{\beta} x^T)}{1 + \exp(\boldsymbol{\beta} x^T)}$$

Here, $\boldsymbol{\beta}$ is a vector of model parameters, and $\boldsymbol{x}$ remains a vector of hydroclimatic independent variables. With some manipulation, equation 5 can be written as:

$$\log\left(\frac{\Pr(I = 1|x)}{1 - \Pr(I = 1|x)}\right) = \boldsymbol{\beta} x = \beta_0 + \beta_1 x_1 + \cdots + \beta_n x_n$$

where $x_a$ and $\beta_a$ are the $a^{th}$ explanatory variable and corresponding coefficient, respectively, and $\beta_0$ is a constant. The left-hand side of equation 6 is the logarithm of the odds-ratio, or the logit. The sample estimates of $\boldsymbol{\beta}, \widehat{\boldsymbol{\beta}}$, can be estimated using maximum likelihood methods that are standard in most software programs. Importantly, such programs also provide estimates of the statistical significance of these estimates, as well as statistics to determine whether the addition of more factors in $\boldsymbol{x}$ is justified. The significance can also be confirmed using bootstrapping, as described below.

By fitting a logistic regression model to historical data, we can determine what hydroclimatic factors are most predictive of the occurrence of ice jam floods. In systems for which there is a small true probability of a flood, $p_{flood}$, one would have to wait a long time to acquire a dataset containing enough floods to accurately estimate $\hat{p}_{flood}$ and the effect of hydroclimatic covariates on $p_{flood}$, $\hat{\boldsymbol{\beta}}$. In such cases the number of flood-years is low, and given the short record lengths the standard logistic regression likelihood approach tends to bias estimates of $\boldsymbol{\beta}$ to be large in absolute value, and thus provides overly certain predictions of $Pr(I = 1|\boldsymbol{x})$ into the future. One method to reduce this bias is to maximize a penalized likelihood function, as proposed by Firth [1993]. Firth regression adds to the standard likelihood equation a penalty term that accounts for the statistical leverage of the data points and is proportional to their information content. The term is equivalent to Jeffrey's prior, so the resulting maximum likelihood estimate, $\hat{\boldsymbol{\beta}}$, corresponds to the Bayesian posterior mode. Conveniently, Firth's method can be applied to small and large samples, where large samples will be less affected by the penalty. Below we implement Firth's regression, along with several other small sample size statistical metrics whose corrections tend to vanish to 0 when using larger datasets.

## 3.1 Regression Model Selection

A forward step-wise approach is taken to model selection (James *et al.*, 2017). In this approach, parameters are added to the model sequentially. When the addition of parameters increases the complexity of the model while not providing sufficient improvement, as measured by some statistic, the selection process terminates in favor of the simpler model. At each step, the explanatory variable that most improves the current model is added. For example, suppose one had five potential explanatory variables: $X_a: a = \{1, 2, \ldots, 5\}$. One begins with the constant model that has no $X$ variables, then determines if the addition of any single variable improves the model fit. The performance of all models including a single explanatory variable are computed and all models that are determined to be better than the constant model are compared to each other. The best performing of these models is selected; for illustration we'll assume $X_2$ was selected. It is then determined if the addition of another variable is an improvement over the model with a constant and $X_2$. This process continues until no improvements can be made.

This forward step-wise approach was used to fit models of varying complexity to the 55-years of data for which freeze-up elevation is available, excluding 1968-1971 when the Williston Reservoir was filling. For each model, the fitted coefficients and the second-order estimate of the Akaike Information Criterion (AICc, small sample size correction) are reported in Tables S.1-S.3, and the best model at each step of the algorithm is reported in Table 2. AICc is one useful metric for model selection, where the model with the lower AICc is generally the best model. AICc penalizes models with more parameters (similar to an adjusted $R^2$ in linear regression). Figure 4 plots the AICc for the best models versus the number of parameters included. The curve flattening or increasing indicates that additional parameters are not providing sufficient improvement to justify the added complexity. We note that the improvement in AICc from two explanatory variables (and a constant) to three is small, and using AICc alone it is not clear that the improvement is significant (note that AICc is itself a statistic computed with error). Looking at the coefficient estimates, $\widehat{\boldsymbol{\beta}}$, a three-parameter model cannot be justified because the additional effects are small in absolute value and are not statistically significant.

Freeze-up elevation enters the model selection process in step 4, although it is not statistically significant and it has a small effect; other factors provide more predictive power. For comparison sake, the AICc for the best model containing freeze-up elevation is also plotted in Figure 4. We see that models containing freeze-up are generally not competitive with other models of the same complexity. Because there is not compelling evidence that freeze-up elevation is a leading predictor of ice jam flood occurrence, we set it aside from consideration and focus the analysis on the best model with two explanatory variables that is supported by the data. Though there are longer records available of these variables, we restrict our analysis to the last 55 years because noteworthy uncertainty exists about the magnitude of older historical floods due to the absence of water level monitoring stations and satellite photography.

*Table 2: Best models for forward step-wise data exploration for 55 years with all explanatory variables. \* indicates significant coefficients at the 5% level. "HH Nov. Flows" is the average monthly flow on the Peace River at Hundsons Hope the November prior to breakup.*

| Number of Explanatory Variables | Best Model | | | Best Model including Freeze-up | | |
|---|---|---|---|---|---|---|
| | Variables | $\hat{\beta}$ | AICc | Variables | $\hat{\beta}$ | AICc |
| 0 | Constant* | -1.86 | 42.17 | Constant* | -1.86 | 42.17 |
| 1 | Constant* <br> GP Precip.* | -3.04 <br> 1.99 | 26.90 | Constant* <br> Freeze-up | -2.06 <br> -0.80 | 38.11 |
| 2 | Constant* <br> GP Precip.* <br> Fort Verm. DDF* | -4.84 <br> 2.37 <br> -1.68 | 22.24 | Constant* <br> Freeze-up <br> GP Precip.* | -3.17 <br> -0.70 <br> 1.86 | 25.93 |
| 3 | Constant* <br> GP. Precip.* <br> Fort Verm. DDF* <br> HH Nov. Flows | -5.46 <br> 2.84 <br> -2.26 <br> 0.66 | 22.23 | Constant* <br> Freeze-up <br> GP Precip.* <br> Fort Verm. DDF | -4.40 <br> -0.30 <br> 2.10 <br> -1.42 | 23.65 |
| 4 (same model) | Constant* <br> GP. Precip.* <br> Fort Verm. DDF <br> HH Nov. Flows <br> Freeze-up | -4.95 <br> 2.56 <br> -1.97 <br> 0.57 <br> -0.19 | 23.96 | Constant* <br> Freeze-up <br> GP Precip.* <br> Fort Verm. DDF <br> HH Nov. Flows | -4.95 <br> -0.19 <br> 2.56 <br> -1.97 <br> 0.57 | 23.96 |

The best model in Table 2 supports our physical understanding of the system, and exploratory data analysis suggests it is the combination of high winter precipitation (high discharge potential) and cold winters (high ice resistance) that is important. Ignoring the important role of winter temperature can create surprising results when applied to climate projections, to be discussed at length in the next section. By ignoring temperature, the model containing precipitation only would result in a projected increase in ice jam floods under extreme RCP8.5 scenarios, which is suspect.

When the combined effects of temperature and precipitation are considered, we see a decrease in ice jam flood frequency associated with a warming (and apparently wetter) climate.

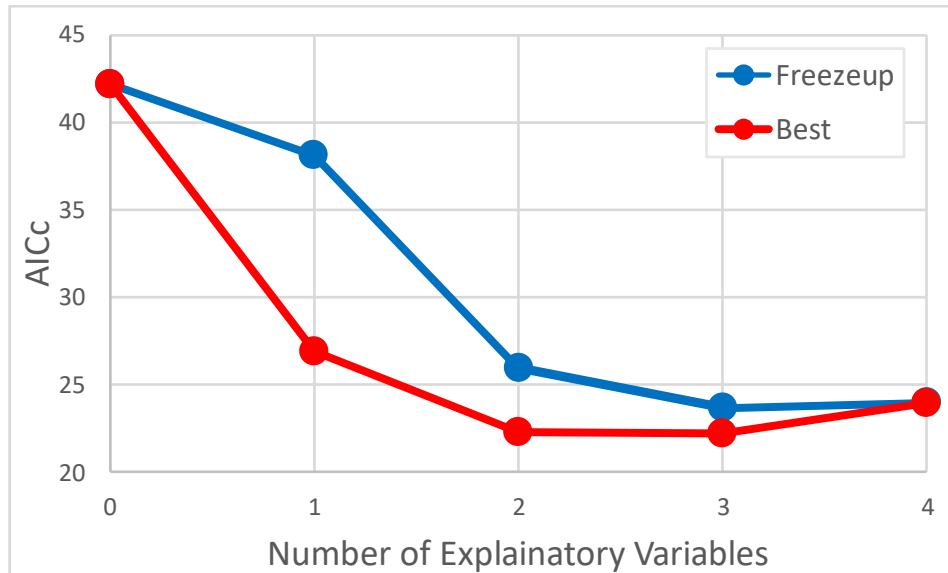

*Figure 4: AICc for the best model (red) and the AICc for the best model including freeze-up elevation (blue) at each stage of the step-forward model selection process*

In addition to the best model in Table 2, we consider several combinations of those two factors:

1. The product of winter precipitation and temperatures (interaction term)
2. The first principle component of the standardized winter precipitation and temperatures

The first combination is simply the product of the winter precipitation and the Fort Vermillion DDF. This product will always be negative, with higher magnitude for winters that are both colder and wetter than average, conditions our mechanistic understanding suggest should be conducive to the formation of ice jam floods.

The second combination is the first principle component of the standardized precipitation and temperature data. Principle component analysis is a rotation (through linear combinations) of the original data to orthogonal axes (Wilks, 2019). The first principle component is the direction of maximum variability in a data set. This is particularly helpful when explanatory variables in a regression are correlated, which can present a problem when interpreting the meaning of the fitted model parameters. There is moderate correlation between the winter precipitation and temperature (Pearson's $r \approx -0.60$), so that some of the information in the temperature record is redundant. The first principle component of the precipitation and temperature data describes about 80% of the variability in the full dataset. Table 3 reports the AICc for each of the combined precipitation-temperature metrics, and the best two parameter model from Table 2.

*Table 3: Comparison of the best bivariate model and various models containing combinations of precipitation and temperature.*

| Variables | $\hat{\beta}$ | AICc |
|---|---|---|
| Constant* | -1.86 | 42.17 |
| Constant* | -4.84 | 22.24 |
| GP Prec.* | 2.37 | |
| Fort Verm. DDF* | -1.68 | |
| Constant* | -3.94 | 22.76 |
| GP Prec.* | 1.80 | |
| Fort Verm. DDF | -0.02 | |
| Interaction | -1.40 | |
| Constant* | -2.56 | 37.25 |
| Interaction* | -0.84 | |
| Constant* | -4.64 | 21.18 |
| 1st PC* | -2.76 | |

In Table 3 we see a slight improvement in AICc when using the first principle component of winter precipitation and temperature compared to the winter precipitation and temperatures directly. However, this improvement is very slight, and in light of the error inherent in estimates of AICc, not strong evidence for the choice of one model over the other. Furthermore, we see very little practical difference between the two models' estimated probability of ice jam floods over the historical period (see Figure S.1). Thus, we prefer the model containing precipitation and temperature over that containing their first principle components, as we deem it simpler from a conceptual standpoint while providing virtually identical answers with the available data.

## 3.2 Model Uncertainty

All of the models reported in the prior section contain error, in part because they were fit to the historical record that is relatively short, as ice jam floods are infrequent. The model coefficients are themselves random variables, and the values reported in the prior section (and SI) are just point estimates, in this case estimated using the Firth [1993] maximum likelihood estimation. It is important to report the uncertainty in our estimates in order to honestly convey the confidence in our predictions. Most logistic regression software packages provide useful information, such as the standard error of the coefficients, which are themselves asymptotic estimates based on model assumptions. To be sure uncertainty estimates are robust to those assumptions being wrong, we apply a parametric bootstrap to estimate confidence intervals for our predicted probability of an ice jam flood and corresponding model parameters (Kéry and Royle, 2016; MacKenzie *et al.*, 2018).

The non-parametric booststrap is a commonly applied, computer-intensive approach to estimate the uncertainty of statistics whose true sampling distributions are unknown or complicated to compute (Efron, 1979). It works by creating many new, random samples by repeatedly re-sampling the original data with replacement. The statistic of interest is then computed on each of the samples, resulting in a distribution of the statistic values from which uncertainty information can be extracted. The parametric bootstrap is similar but begins with the assumption that the underlying distribution is known, but that the parameters of that distribution are unknown (Kéry and Royle, 2016). The independent variables are fixed, and new samples of the dependent variable timeseries are generated by randomly simulating from the assumed distribution with parameters estimated from the data. In a straight-forward application of the parametric bootstrap to our example, each re-sample includes the exact historical hydroclimatic record, but with random ice jam flood timeseries generated from a Bernoulli random variate with $p_{flood}$ equal to the predicted probability from the best model in the previous section. This process is repeated $B$ times (say 1,000), with a new logistic regression model fit to each of the $B$ samples, resulting in $B$ estimates of the model parameters from which uncertainty information can be extracted.

*Table 4: Bootstrapped 95% CI for fitted model parameters, and the parameter values as estimated by Firth's regression using the historical data.*

| Parameter | $\hat{\beta}$ | 95% CI |
|---|---|---|
| Constant | -4.84 | [-11.34, -2.50] |
| GP Precip. | 2.37 | [0.85, 5.42] |
| FC DDF | -1.68 | [-4.70, -0.30] |

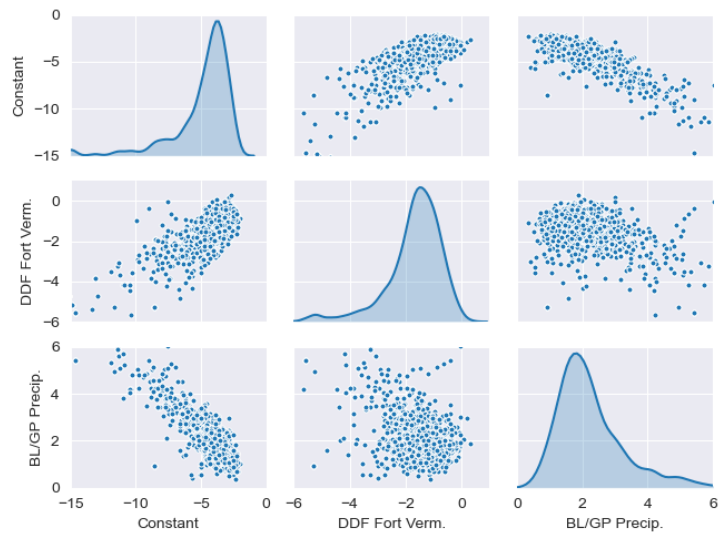

*Figure 5: Empirical distribution of logistic regression parameters from the parametric bootstrap. Axes have been limited for clarity, excluding 2% percent of boot strap fits.*

Figure 5 plots the individual and joint distribution of coefficients of the fitted model ($\hat{\beta}_o$, $\hat{\beta}_{DDF}$, $\hat{\beta}_{prec}$), and Table 4 reports the 95% CI for the three fitted parameters. We see that the 95% confidence intervals (CI) do not include zero for any of the parameters, which is consistent with the p-values estimated by Firth (1993) regression (i.e. all parameters of the best model are significant at the 5% level, see Table 3). Yet, there is large uncertainty as can be observed by the wide 95% CI in Table 4 and the wide distributions in Figure 5. This means that, while we are quite certain that colder winters and higher winter precipitation increase the chances of an ice jam flood, it is difficult to estimate by precisely how much they do. Each point in the scatter plots in Figure 5 represents a fitted model to an alternative history that might have occurred but did not by chance. In the next section we will use these models to estimate potential changes to PAD ice jam flood frequency under climate change.

## 3.3 Future Flood Frequency in the PAD

Our first research question asked what are the key factors that cause PAD ice jam floods, and our analysis identified (1) Fort Vermillion DDF and (2) Beaverlodge/Grande Prairie winter precipitation using logistic regression. This model is consistent with our physical understanding of the system, and we find that large ice jam floods have been rarely observed when winter temperatures are warmer or winter precipitation is lower than the historical average (see Section 2.2 and Figure 3). This is true over the last 100 years, though the magnitude of floods over the historical period is itself uncertain. Given this relationship, our second research question asks how future climate change may impact the occurrence of large PAD ice jam floods and typical waiting times between their occurrence.

Answering this question is challenging because the future is shrouded by various types of uncertainty. In the long-run and at the large-scale there is deep uncertainty about future emissions and the climate's response to those emissions (Lempert, 2002; Lamontagne *et al.*, 2018). At a finer scale there is epistemic uncertainty about the exact relationship between climate forcing and ice jam flood probability (see Section 3.2). There is also aleatoric (random) uncertainty about future weather, and whether an ice jam flood may occur or not in any year. Indeed, the 2020 Peace River breakup is an example for which conditions for an ice jam flood seemed good, but no flood occurred due to many small factors that would be difficult to observe or predict. We incorporate these uncertainties into a unified simulation framework, centered on the logistic regression model from the previous section (see Figure 6), which can shed light on future trends in PAD ice jam floods.

To represent the deeply uncertain future emissions trajectories and the subsequent climate response, the framework begins with two representative concentration pathways (RCPs, van Vuuren *et al.* (2011)) and six climate models (HadGEM2, ACCESS, CanESM, CCSM4, CNRM-CM5, and MPI-ESM-LR) that are downscaled to provide annual estimates of our predictors, ***X***: winter DDF at Fort Vermillion and winter precipitation at Grand Prairie, respectively (Cannon, Sobie and Murdock, 2015; Werner and Cannon, 2016). To represent epistemic uncertainty in the relationship between these climate forcings and the occurrence of ice jam floods, random logistic regression models from the parametric bootstrap described in the previous section (Figure 5) are drawn and forced by the downscaled GCM projections for 2020-2100. Taken in aggregate, the results from the sampled models provide trends of the probability of ice jam flood occurrence for each GCM-RCP scenario. From these ensembles we compute metrics of interest for each scenario, such as cumulative $21^{st}$ century floods and median wait times between floods. The experimental set-up is illustrated in Figure 6.

Figure 7 shows the 50% confidence corridors (interquartile range) for the probability of an ice jam flood from 1982 to 2100 for each of the GCMs considered (one per panel), and each climate scenario (RCP 4.5 in blue, RCP 8.5 in green). We see that over the historical period there was a reduction in the estimated probability of ice jam floods from the 1980s into the early 1990s, after which the estimated probability remained fairly constant between 0.15 and 0.05 through 2020. After the historical period, the probability corridors begin to separate by both GCM (panel to panel) and RCP (by color). Through the middle of the $21^{st}$ century, some models such as CNRM-CM5 and to a lesser extent CanESM2 agree across RCPs, despite substantial differences in the underlying emissions trajectory. Other models show order of magnitude deviations between RCPs, and the deviation between climate scenarios grows in almost all models over time. For instance, HadGEM2-ES and CCSM4 estimate the average annual probability of a large PAD ice jam flood at the end of the century to be about 2.2% and 1.2%, respectively. Despite differences between the models and emissions scenarios about precise flood probabilities and their uncertainty, all projections agree that the probability of ice jam floods will decline during the 21st century, likely by orders of magnitude. However, as shown in 95% confidence corridors in Figure 8, most models' extremes show some chance that the probability at the end of the century will be unchanged from conditions today. Essentially, winters are expected to be insufficiently cold to build substantial river ice and snow packs are expected to be reduced, making ice jam floods unlikely.

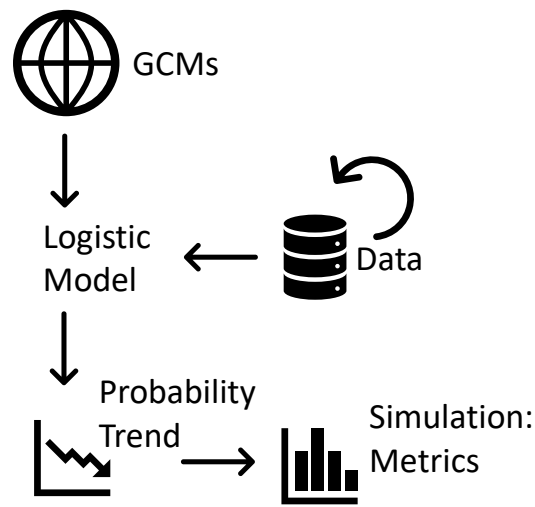

*Figure 6:* Simulation framework for future PAD ice jam floods. The framework begins with downscaled GCM projections of winter temperatures and precipitation from six GCMs and two RCPs, which force a randomly generated logistic model generated by bootstrapping the original data. This provides a time series of conditional probabilities of ice jam floods, which are used in a Monte Carlo simulation to generate ensembles of metrics discussed in the text.

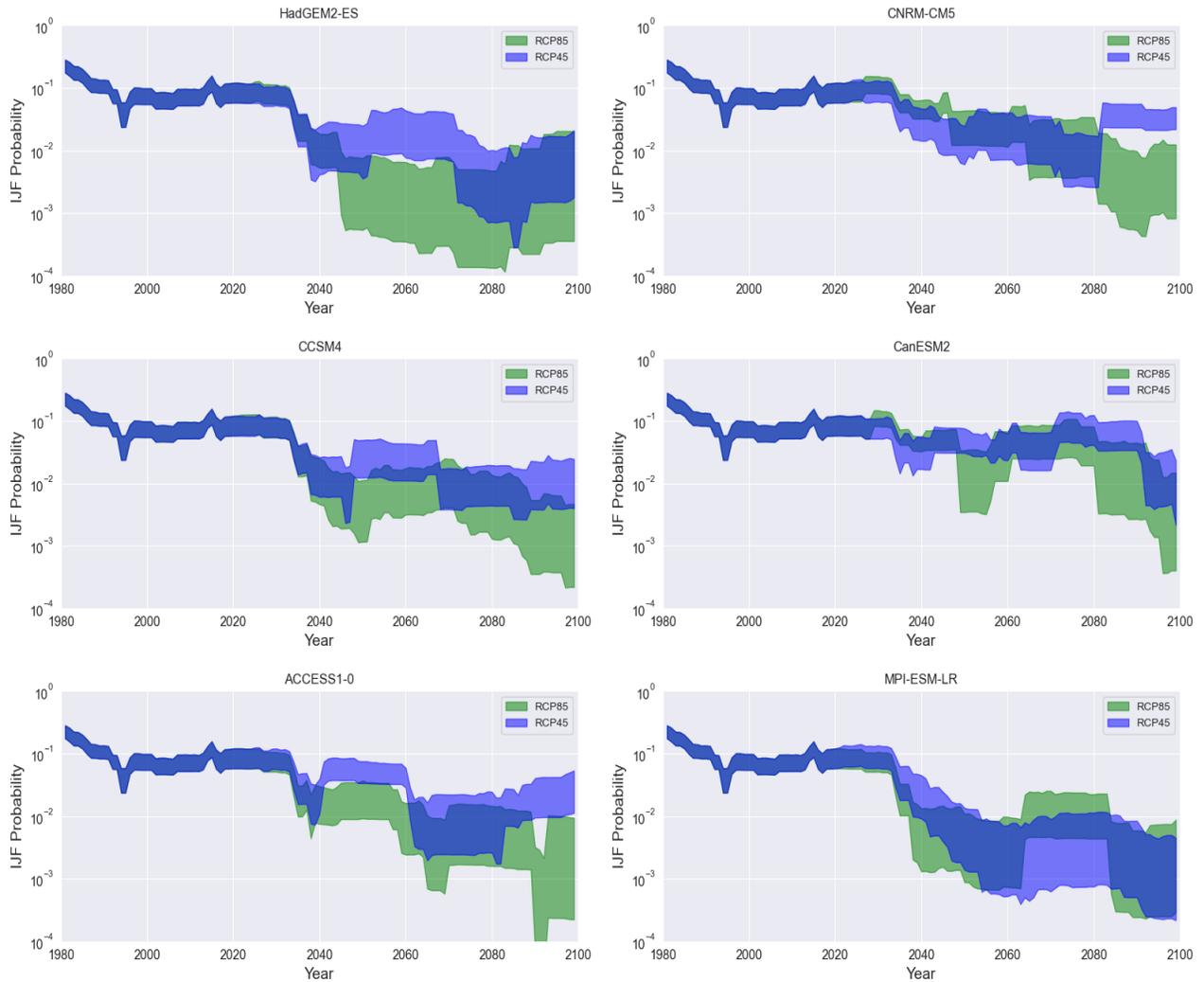

Figure 7: 50% confidence corridors (interquartile range) for the probability of a large PAD ice jam flood over the historical period (1962-2020) and under twelve climate projections (2020-2100). A 20-year moving average in real space is used for plotting, so 1982 is the first plotted year. Each panel shows a different GCM, with the blue corridor representing RCP45 and the green corridor representing RCP85. The y-axis shows the probability of an ice jam flood in log-space, so a reduction by 1 tick mark corresponds to an order of magnitude reduction in probability.

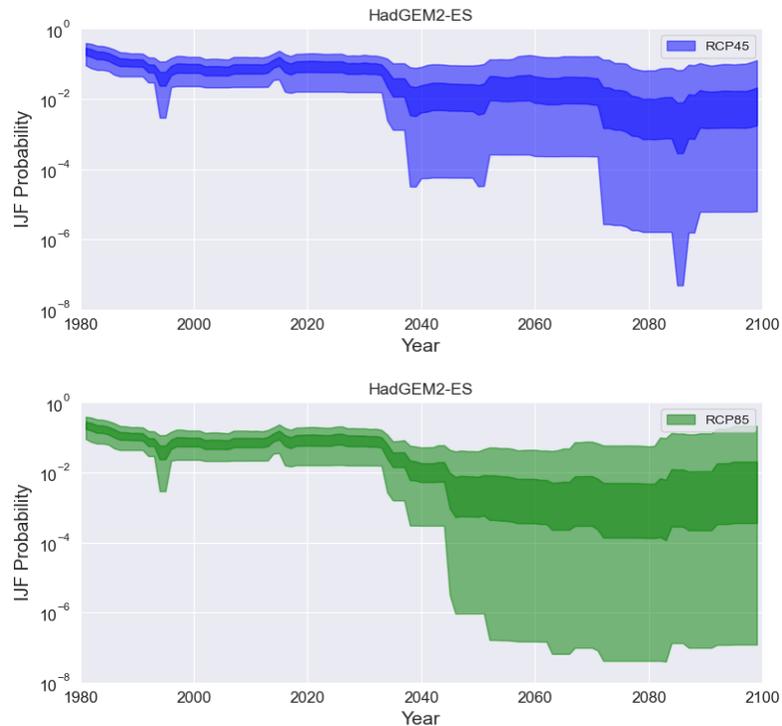

*Figure 8:* 9*5% confidence corridors for* the HadGEM2-ES GCM in *Figure 7. The 50% confidence corridor is displayed in a darker shade.*

Of particular concern for the ecology of the PAD are prolonged periods without large ice jam floods, or alternatively changes to the distribution of wait times between ice jam floods. Figure 9 plots the 50% confidence corridors for the ice jam flood return period for each GCM (panels) and RCP (colors) scenario. From the 1980s to the mid-1990s the return period increases, then stabilizes around 11 years. In most projections, this return period persists through about the 2030s before increasing. The rate and amount of the increase again depends heavily on the GCM and the RCP. Under RCP45 the CanESM2 model shows that return periods remain largely unchanged through much of the 21$^{st}$ century, while for RCP45 MPI-ESM-LR the return period climbs to about 140 years by 2050. Despite this disagreement, across the scenarios and GCMs, return periods increase by two-to-three orders of magnitude by the end of the century, with RCP85 generally seeing greater increases. Return periods between 100- and 1000-years for large ice jam floods are common in our simulation for high warming (RCP85 scenarios). This makes good physical sense given our mechanistic understanding of the system, though it would likely mark a significant change to the PAD's current hydroecology. However, the interpretation of the return period is difficult in a non-stationary world, such as these climate projections provide, wherein the probability of some extreme event is changing systematically over time (Read and Vogel, 2015). A more intuitive metric is the waiting time until the next flood. That is, supposing you are in 2030 in some climate projection, how long must you wait to observe the next ice jam flood? Of course, this waiting time is a random variable, with its own distribution, and one may be interested in different aspects of the distribution depending on application (e.g. median waiting time vs. 95$^{th}$ percentile waiting time).

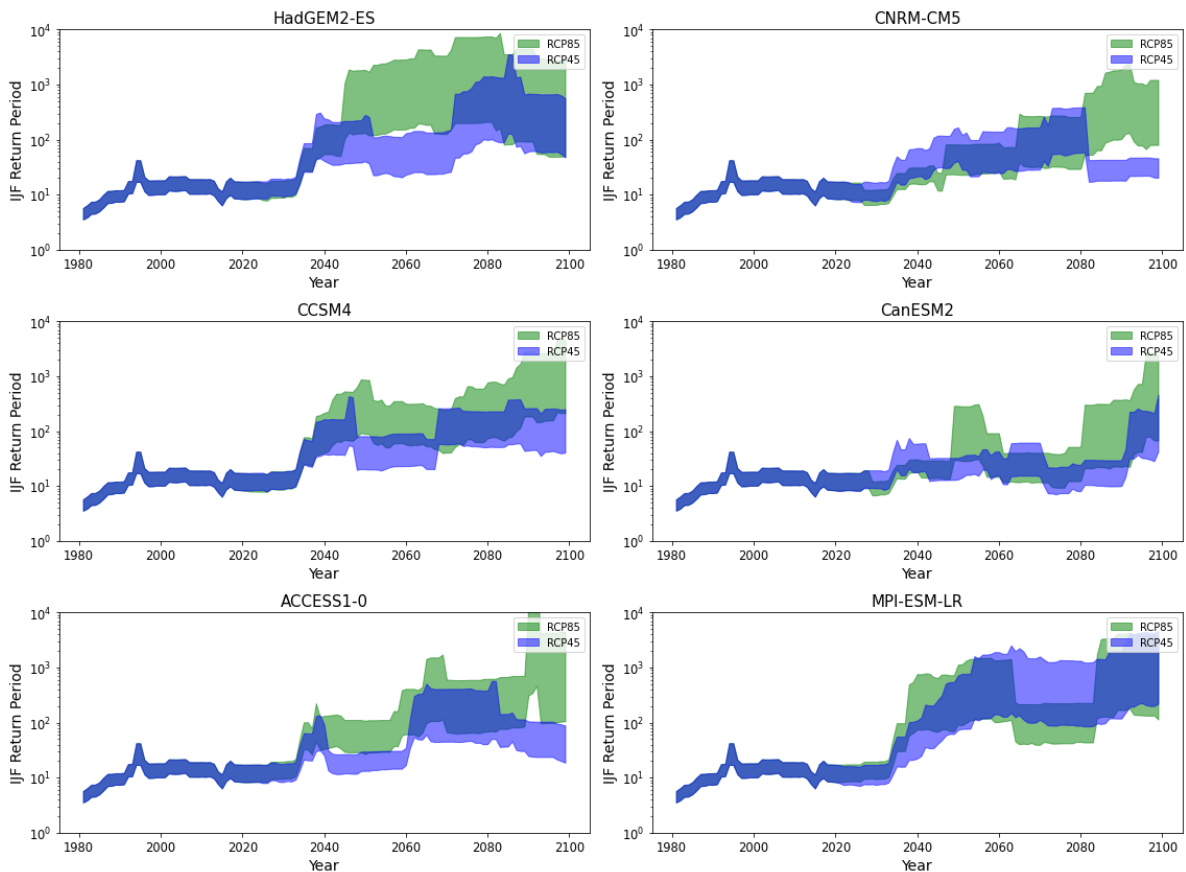

*Figure 9: 50% corridors for the instantaneous return period for large PAD ice jam flood over the historical period (1980-2020) and under twelve climate projections (2020-2100). Each panel shows a different GCM, with the blue corridor representing RCP 4.5 and the green corridor representing RCP 8.5.*

*Table 5: Median wait-times (years) for a large PAD ice jam flood in 2030 and 2050 for six GCMs and two RCPs. Survival analysis is used (Klein and Kleinbaum, 2011; Cameron Davidson-Pilon* et al.*, 2021) for cases when wait times extend beyond the end of the simulation.*

| RCP | GCM | 2030 | 2050 |
|---|---|---|---|
| 4.5 | HadGEM2-ES | 57 | 58 |
| | ACCESS1-0 | 11 | 47 |
| | CanESM2 | 13 | 11 |
| | CCSM4 | 18 | 62 |
| | CNRM-CM5 | 42 | 32 |
| | MPI-ESM-LR | 204 | 178 |
| 8.5 | HadGEM2-ES | 142 | 126 |
| | ACCESS1-0 | 71 | 98 |
| | CanESM2 | 27 | 11 |
| | CCSM4 | 108 | 132 |
| | CNRM-CM5 | 31 | 72 |
| | MPI-ESM-LR | 108 | 96 |

For each of the twelve RCP-GCM combinations we simulate 1,000,000 future flood sequences using the framework in Figure 6 (1,000 replicates for each of the 1,000 bootstrapped models). From these 1,000,000 futures we construct the distribution of wait times for each year in the future projections. For brevity, we consider the median wait time in 2030 and 2050, which are reported in Table 5. For RCP45, median wait times for most GCMS are between about 10 and 50 years. ACCESS1-0, CanESM2, and CCSM4 have median wait times that are consistent with recent historical experience, while HadGEM2-ES and CNRM-CM5 have median wait times greater than 40 years, meaning the next ice jam flood does not occur until at least the 2070s in half of the 1,000,000 futures. By the 2050s, median wait times have increased for most models, and many models see no further large PAD ice jam floods for the duration of the 21$^{st}$ century in half of the futures. MPI-ESM-LR's RCP45 realization is an anomaly, with substantially lower mid-to-end-of century ice jam flood probabilities, long return periods, and exceptional wait times. Future work should consider a larger number of MPI-ESM-LR realizations to determine if this is a sampling anomaly or a fundamental difference in the projected climatology. For RCP85, median wait times are substantially longer than RCP45, for most models. After 2030 at least half of futuress see no further 21$^{st}$ century large ice jam floods for HadGEM2-ES, CCSM4, and MPI-ESM-LR. By 2050 the median wait time exceeds the remaining duration of the 21$^{st}$ century for all models except CanESM2. While these wait times are long relative to recent experience, they are not inconsistent with wait-times found in the paleo-record (Timoney *et al.*, 1997; Wolfe *et al.*, 2006; Hugenholtz, Smith and Livingston, 2009; Timoney, 2013).

While there is large variability across the six GCMs and two RCPs concerning the future of large PAD ice jam floods, we can draw three important conclusions from Figure 7, Figure 9, and Table 5. First, the frequency of PAD ice jam floods is likely to decrease as the PAD's climate changes in the coming decades, and this decrease will likely be sizeable relative to recent history. Second, the extent of the change in ice jam frequency and typical wait times depends as much on the choice of GCM as RCP: the choice of GCM model is more impactful in some cases than the assumed inputs to that model. Finally, projections of future PAD ice jam flood frequency are subject to large uncertainty. Ignoring the deep uncertainties around GCM and RCP for a moment, the 50% confidence intervals on the ice jam flood probability are very broad, covering an entire order of magnitude. Thus, while we can robustly say that ice jam floods will decrease in frequency, likely substantially, it is difficult to say by precisely how much without making many assumptions, and all projections should be made humbly.

## 4 Discussion

Large ice jam floods in the PAD are relatively rare events that are caused by a complicated sequence of precursory events, some of which are difficult or impossible to observe. As a consequence of observing only some of the factors that cause large ice jam floods, deterministic predictions of ice jam flood occurrence have limited accuracy, and deterministic modeling of ice jam floods is generally restricted in scope to specific events for which many factors are observed. This suggests that stochastic methods to assess ice jam flood frequency, past and future, may provide important insight for the most general case of observing few factors. Those methods are, however, limited by the length of the reliable flood and climatic records, so any statistical result must be understood in the context of its uncertainty. This work merged insights from decades of mechanistic research on ice jam physics with appropriate statistical methods to derive new insights about ice jam flood frequency. Specifically, we utilized a recent taxonomy of the necessary conditions for ice jam flood formation to inform the construction of a statistical model, which is fit using proper techniques for short records to assess the strength of the relationships suggested by physical theory.

In this analysis we find that winter temperatures upstream of the PAD (Fort Vermillion DDF) and winter precipitation at Grande Prairie are most predictive of large ice jam floods. The predictive power of statistical models that contain other factors that have been proposed in the literature, such as freeze-up elevation, the rapidity of the spring thaw, and the previous November or December flows, are smaller, with parameter values that are generally not significant and thus are not supported by the data (see Tables S.1-S.3). Based on (Jasek, 2019c)'s taxonomy, it is surprising that colder winters at Fort Vermillion (and presumably thicker ice upstream of the PAD) are more predictive of large ice jam floods than colder winters at Fort Chipewayan (and presumably thicker ice in the PAD). However, there is a very strong correlation between winter temperatures in the two locations ($r = 0.93$) so it is unlikely that our model distinguishes between ice resistance in and upstream of the PAD, but instead indicates that winters must be sufficiently cold to form major ice resistance in the area. The effect and timing of snowfall, snowdrift, and snow settlement on the ice cover and their corresponding influence on thermal ice growth is not available, which also contributes uncertainty in determining whether DDF at Fort Vermilion or Fort Chipewyan are more predictive.

Freeze-up elevation is a proxy for bank resistance to the ice sheet linked to the geometry of the ice sheet getting around bends and through narrow openings, while DDF is one of many proxies for the ice resistance to fracture at the time of breakup. In the latter case, the ice sheet does not have to be narrow enough to get around bends for a dynamic breakup to continue downstream but just weak enough to crack into pieces that are small enough to get around bends and narrows. Cracks can form quickly from point loads from islands and riverbank planform irregularities as driving forces (water flow and gravity) push the ice sheet downstream. The Peace River is a wider river than most, so it is likely that ice resistance rather than bank resistance plays a more dominant role because the wider the ice sheet compared to its thickness, the more likely weak spots will allow crack propagation from point loads.

Another important consideration is the ice bending strength. This is what causes the arrests of dynamic breakups that lead to ice jam formation and flooding. Strong bending strength is needed to allow resistance to further breakage from the buoyancy of the accumulation of ice rubble under the ice cover at the breakup front (Andres, Jasek and Fonstad, 2005). If the ice cover is strong enough in bending, large quantities of ice rubble underneath can reach the river bed providing anchor points from the river bed causing the arrest of the dynamic breakup leading to ice jam formation and flooding. Freeze-up elevation does not play as direct a role to thickness and strength of the ice sheet as does DDF in this mechanism.

In absolute terms, none of the models have strong predictive power, and this reflects the complexities of the real system and the limited ability to observe precursory events and the factors that cause them. Even if all observable conditions are perfect, a large ice jam flood may or may not form due to these unobserved factors. This highlights that with the available data, no statistical model of PAD ice jam frequency can speak authoritatively about what would or would not have happened under hypothetical conditions: the best we can do is speak to what is more or less likely.

The short reliable record length in the PAD limits our ability to estimate the parameters of any statistical model. In practical terms, this means that we cannot be precisely sure how much winter temperatures or precipitation increase the chances of a large ice jam flood. Firth's logistic regression provides estimates of the significance of each parameter in our model using a p-value, which is the probability (given the model assumptions) that as extreme a parameter value arose from chance. The parameters of the best model have p-values below 0.016. To test the significance of our model parameters making fewer assumptions, we employed a parametric bootstrap that simulates many alternative histories, fits a model to each, and provides a sense for the joint sampling distribution of the fitted model parameters. The results agree strongly with the standard logistic regression p-values, which provides further support for the choice of predictor variables (winter temperature and precipitation). It should be noted that other uncertainties, such as structural model uncertainties and uncertainties about the data (particularly the magnitudes of floods in years prior to our selected dataset), are not considered here but should be considered in future work to further refine the model and ideally improve predictive power.

The future of PAD ice jam flood frequency is shrouded by deep uncertainty concerning future greenhouse gas emissions and the climate's response to those emissions. We try to assess future ice jam flood frequency under these uncertainties by using downscaled climate projections for two emissions trajectories and six GCMs as forcing for our logistic regression model of ice jam flood occurrence. We find that despite some disagreement between GCMs and RCP scenarios, most agree that the PAD will see order of magnitude decreases in future ice jam flood probability and that wait times between ice jam floods will increase substantially. Although the projected changes in ice-jam flood frequency are sizeable and would imply substantial hydroecological changes for the PAD, the changes are not unprecedented given the decadal-scale intervals without large floods reconstructed from oxbow lake sediment records (Wolfe *et al.*, 2006). Assessing the likelihood of future emissions scenarios, or the reliability of the various GCMs over specific spatial or temporal domains is beyond the scope of this work. However, it should be noted that in many cases the chosen GCM is a stronger determinant of future ice jam flood frequency than the emissions trajectory used to force the GCM.

The frequentist approach to model fitting, the informal use of expert information, and the treatment of uncertainty using bootstrapping represent a first step in the stochastic modeling of PAD ice jam flood frequency. Subsequent work should incorporate prior (expert) information more formally in the fitting procedure using Bayesian modeling techniques. More historical information (extending to the early 20$^{th}$ century) could also be incorporated in the modeling procedure by explicitly accounting for the large uncertainty concerning the magnitude of historical floods, which is a topic of debate in the community that will be explored in future work.

## 5 Conclusions

PAD ice jam flood frequency and its dependence on various factors is challenging to estimate and accurately predict due to unobservable factors and limited data. Any claims of predictive power should be presented and interpreted in light of the great uncertainty inherent in the problem. With sound statistical techniques supported by expert understanding of the physical factors of ice jam flood occurrences, it is possible to provide insight into which factors more likely do predict Peace River ice jam floods (winter temperatures and precipitation), which factors more likely do not (like freeze-up elevation and early winter flows), and the uncertainty in our estimates of those relationships. If current trends in greenhouse gas emissions persist, the PAD will likely see a substantial decline in ice jam flood frequency, and the waiting time between floods will grow accordingly, though by precisely how much depends on future emissions and the climate's response to those emissions. Still, all models and scenarios agree that climate change will likely push ice jam flood frequency below historical experience, which could change the PAD's ecology in ways that are difficult to anticipate.  This work establishes an introduction to stochastic modeling of ice jam flood frequency in the PAD that accounts for climactic factors that likely influence physical flood causes.  Future work should build upon this study by making more formal use of expert knowledge and extend the record by accounting for uncertainty in historical flood data within a Bayesian framework.

## Acknowledgments:

The authors would like to thank Kevin Timoney for his advice concerning the ecological aspects of the PAD, and Georg Jost for his advice on climate scenarios and provision of downscaled climate forcings for the PAD.  This research was supported in part by a grant from BC Hydro.  All data and code used in this analysis can be found in this repository: https://doi.org/10.5281/zenodo.4474826

# 7 Supplement

*Table S. 1:* Univariate Models fit to 55 years for which all hydroclimatic, freeze-up elevation data are available. Parameters significant at the 10% level are bolded. * indicates that the record length is reduced due to 1 (PR) or 5 (PP) years of missing data, and so models are not comparable to 55-year length models. HH, PR, and PP stand for Hudsons Hope, Peace River, and Peace Point respectively, where there are hydrometric stations located on the Peace River.

| Variable Name | $b_1$ | $b_0$ | $AICc$ |
|---|---|---|---|
| Constant | | **-1.86** | 42.17 |
| GP % of avg Precip Nov-Apr | **1.99** | **-3.04** | 26.90 |
| Fort Vermilion DDF | **-1.70** | **-2.73** | 30.67 |
| PR previous Nov Flows* | -0.50 | **-1.89** | 40.42 |
| Fort Chip. DDF | **-1.85** | **-2.82** | 30.62 |
| Melt Test 8 | -0.51 | **-1.90** | 41.16 |
| Freeze-up Elevation | **-0.80** | **-2.06** | 38.11 |
| HH previous Nov Flows | -0.58 | **-1.95** | 39.84 |
| PP previous Nov Flows* | -0.65 | **-2.03** | 34.96 |
| HH previous Dec Flows | **-0.66** | **-2.00** | 38.45 |
| PP previous Dec Flows* | **-0.82** | **-2.14** | 32.98 |
| PR previous Dec Flows* | **-0.67** | **-1.98** | 38.23 |

*Table S.2:* Bivariate Models including GP Precip. fit to 55 years for which all hydroclimatic, freeze-up elevation data are available. Parameters significant at the 10% level are bolded. * indicates that the record length is reduced due to missing data, and so models are not comparable to 55-year length models. HH, PR, and PP stand for Hudsons Hope, Peace River, and Peace Point respectively, where there are hydrometric stations located on the Peace River.

| Variable Name | $b_2$ | $b_1$ | $b_0$ | $AICc$ |
|---|---|---|---|---|
| Fort Vermilion DDF | **-1.68** | **2.37** | **-4.84** | 22.24 |
| PR previous Nov Flows* | -0.04 | **1.90** | **-2.88** | 27.76 |
| Fort Chip DDF | **-1.53** | **1.78** | **-3.96** | 24.02 |
| Melt Test 8 | -0.43 | **1.88** | **-3.03** | 27.40 |
| Freeze-up Elevation | **-0.70** | **1.86** | **-3.17** | 25.93 |
| HH previous Nov Flows | -0.14 | **1.88** | **-2.92** | 27.57 |
| PP previous Nov Flows* | -0.32 | **2.14** | **-3.40** | 23.31 |
| HH previous Dec Flows | -0.22 | **1.86** | **-2.98** | 27.08 |
| PP previous Dec Flows* | -0.10 | **2.14** | **-3.26** | 23.44 |
| PR previous Dec Flows* | -0.26 | **1.84** | **-2.94** | 26.94 |

*Table S.3:* Trivariate Models including GP Precip., Fort Vermillion DDF fit to 55 years for which all hydroclimatic, freeze-up elevation data are available. Parameters significant at the 10% level are bolded. * indicates that the record length is reduced due to missing data, and so models are not comparable to 55-year length models. HH, PR, and PP stand for Hudsons Hope, Peace River, and Peace Point respectively, where there are hydrometric stations located on the Peace River.

| Variable Name | $b_3$ | $b_2$ | $b_1$ | $b_0$ | $AICc$ |
|---|---|---|---|---|---|
| PR previous Nov Flows* | 0.86 | **2.92** | **-2.40** | **-5.62** | 21.94 |
| Fort Chip DDF | 0.16 | **2.27** | -1.72 | **-4.62** | 25.48 |
| Melt Test 8 | 0.35 | **2.00** | -1.58 | **4.10** | 24.13 |
| Freeze-up Elevation | -0.30 | **2.10** | -1.42 | **-4.40** | 23.65 |
| HH previous Nov Flows | 0.66 | **2.84** | **-2.26** | **-5.46** | 22.23 |
| PP previous Nov Flows* | 0.08 | **1.86** | -1.32 | **-4.01** | 22.46 |
| HH previous Dec Flows | 0.38 | **2.58** | **-2.01** | **-5.06** | 22.50 |
| PP previous Dec Flows* | 0.49 | **2.52** | **-1.80** | **-4.98** | 21.51 |
| PR previous Dec Flows* | 0.38 | **2.62** | **-2.04** | **-5.13** | 22.62 |

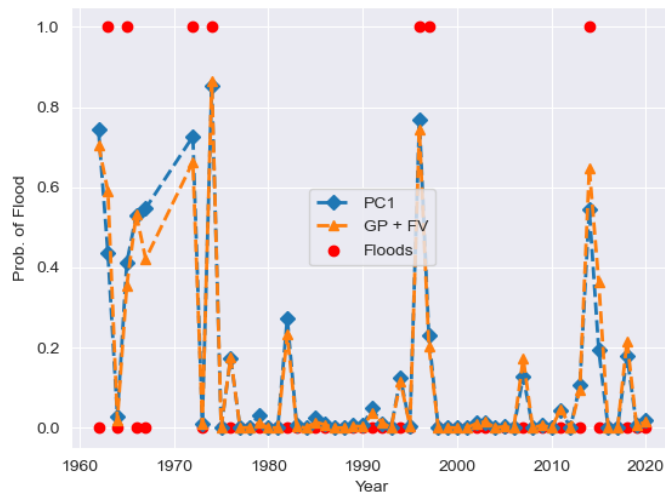

*Figure S.1 Comparison of PC and precipitation & temperature models*